\documentclass[prf,aps,11pts, twocolumn,notitlepage,longbibliography]{revtex4-1}
\usepackage{graphicx}
\usepackage[usenames]{color}
\usepackage{amsmath,amssymb}
\usepackage{gensymb}
\usepackage{natbib}
\usepackage{tabularx}
\usepackage{xcolor}
\usepackage{upgreek}
\usepackage{hyperref}
\usepackage{subfigure}
\usepackage{changes}

\usepackage{upgreek}

\newcommand{\comA}[1]{\textcolor{black}{ #1}}

\usepackage{multirow}

\begin{document}
\title{Microfluidic {In Situ} Measurement of Poisson's Ratio of Hydrogels}

\newcommand{\orcidauthorA}{0000-0000-000-000X} 

\author{Jean Cappello $^{1}$, Vincent d'Herbemont $^{1}$, Anke Lindner $^{1,}$* and Olivia du Roure $^{1}$}


\affiliation{%
$^{1}$ \quad  Laboratoire PMMH-ESPCI Paris, PSL Research University, Sorbonne Universit\'e, Universit\'e de Paris, 10, rue Vauquelin, F-75005, Paris, France}


\begin{abstract}
{Being able to precisely characterize the mechanical properties of soft microparticles is essential for numerous situations from the understanding of the flow of biological fluids to the development of soft micro-robots. Here we present a simple measurement technique for the Poisson's ratio of soft micron-sized hydrogels in the presence of a surrounding liquid. This methods relies on the measurement of the deformation in two orthogonal directions of a rectangular hydrogel slab compressed uni-axially inside a microfluidic channel. Due to the \textit{in situ} character of the method, the sample does not need to be dried, allowing for the measurement of the mechanical properties of swollen hydrogels. Using this method we determine the Poisson's ratio of hydrogel particles composed of polyethylene glycol (PEG) and varying solvents fabricated using a lithography technique. The results demonstrate with high precision the dependence of the hydrogel compressibility on the solvent fraction and character. The method, easy to implement, can be adapted for the measurement of a variety of soft and biological materials.}
\end{abstract}


\maketitle


\section{Introduction}

{Soft materials are found in situations as different as the flow of biological fluids, biomedical devices, micro-fluidic sensors, or soft micro-robotics, in forms such as soft polymeric particles or different types of protein or cell aggregates}. Among soft materials, hydrogels are extremely powerful materials, allowing particles of controlled shapes to be manufactured at the micron scale. 
Their biocompatibility, softness, and easy and rapid fabrication make them perfectly suited for biomedical applications like drug delivery \cite{Nayak2004, Soppimath2005} or tissue engineering \cite{Ambrosio1998}. They are also useful as building blocks for the design of soft composite systems that are able to change shape in response to external stimuli, offering applications in soft robotics, flexible electronics, and biosensor development  \cite{Liu2016}.
Being able to precisely characterize the mechanical properties \comA{of soft microparticles} is essential when designing such applications, but faces specific problems due to their small scale, the dependence of their properties on the surroundings, and their ability to evolve over time.

The static linear mechanical properties of an isotropic elastic material are fully determined by two quantities:  Young's modulus ($E$) and Poisson's ratio ($\nu$). In a compressive test,  Young's modulus links stress and strain, whereas Poisson's ratio quantifies the expansion of a material in directions perpendicular to the direction of compression.  Poisson's ratio varies, for an isotropic material, from 0.5 (incompressible) to -1.  Other quantities such as the bulk modulus ($K$), the shear modulus ($G$), or Lam\'e{'s}
~first coefficient ($\lambda$) correspond to different combinations of these two quantities  \cite{Crandall1978}.

Different techniques have been developed to measure  Poisson's ratio. The most straightforward method relies on the measurement of the strain in two orthogonal directions during a compressive test.  Another method is to measure two different moduli,  the bulk modulus $K$ and Young's modulus $E$, and to obtain Poisson's ratio using their relationship $K = E/(3(1-2\nu))$. Classical methods usually require a macroscopic sample, and thus a significant volume of material, which can be limiting for microscopic soft particles. Additionally, in these classical methods the sample often needs to be fixed onto the testing apparatus and thus requires the sample to be dried \cite{Li1993}. For the soft materials considered here, drying may have a strong impact on the microstructure and thus on the material properties. In addition, soft hydrogels may have internal timescales due to poroelasticity and/or viscoelastic effects, which may influence their mechanical responses depending on the frequency or time scale of the chosen technique.

Some methods have been proposed to overcome these limitations. Wyss 
{et al.} \cite{Wyss2010} developed a  simple method using microfluidics to measure the Poisson ratio of a soft particle relying on the independent measurement of the bulk  and Young's modulus. However, the experimental errors of each of these measurements, even more pronounced for soft materials, limit the accuracy of the determination of Poisson's ratio, which only varies over a small range.  
Other techniques, such as micro-pipette aspiration experiments \cite{Li2009, Boudou2006} or atomic force microscopy (AFM) \cite{Hurley2007}, provide high-accuracy measurements of Poisson's ratio, but only locally probe the surface of the sample and do not provide bulk measurements. Acoustic waves can also be used to perform dynamic mechanical tests  \cite{Evans1999}.  Usually the wave frequency varies from 0.5 to 5.0 MHz, and the measured elastic constants might be dependent on the frequency and differ from a static compression test. Finally, techniques such as small-angle X-ray scattering \cite{Moram2007} also enable the measurement of Poisson's ratio, but require costly infrastructures.

\begin{figure*} [ht]
\centering
\includegraphics[width=0.5 \textwidth]{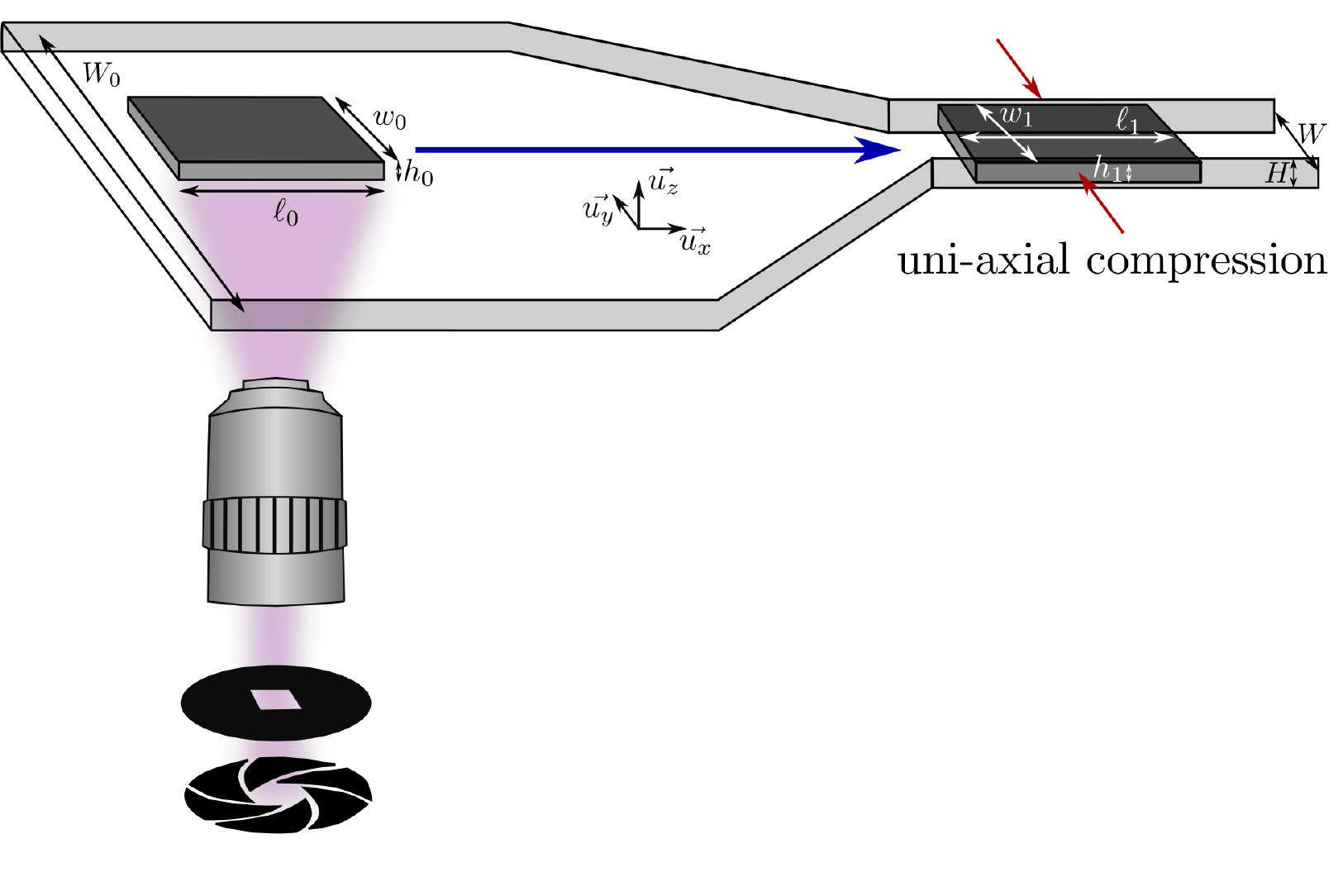}
\caption{
{In situ} {fabrication of a particle using a microscope-based} projection lithography technique. UV light is projected into a flat channel through a shutter and a rectangular mask resulting in the fabrication of a hydrogel particle surrounded by uncured solution. The slab is then pushed into a constriction by applying an external flow. When the particle has completely entered the narrow region of the channel the flow is stopped and the particle experiences uni-axial compression.} 
\label{fig:schema_1}
\end{figure*}

In this article, we propose a simple and direct method to measure {Poisson's ratio} of soft hydrogels at long times directly inside a microchannel.  We use a microfluidic setup where hydrogel particles are fabricated and subsequently transported in a channel whose width is smaller than the particle width (see Figure~\ref{fig:schema_1}). After the full entry of the particle into the narrow channel, the external flow is stopped and the particle is solely submitted to an uniaxial compression by the lateral walls. This geometry is well adapted to determine Poisson's ratio by measuring at equilibrium the deformation of the particle in the direction of the uniaxial compression as well as in the perpendicular direction. {As this method is performed 
{in situ}, no drying steps are required. This ensures that the structure of the material stays unmodified}. 
Temporal observations of the hydrogel deformation allow for access to the time dependence of the process and to unequivocally determine the equilibrium state reached at long timescales. After detailing the method and its limitations we provide accurate measurements of {Poisson's ratio} of polyethylene glycol diacrylate (PEGDA) hydrogels fabricated in different solvents. 

\section{Materials and Methods}

 \subsection{Channel and Particle Fabrication}
PDMS (polydimethylsiloxane, Sylgard 184, DOWSIL) micro-channels were fabricated using traditional soft lithography techniques. The channels were bound to a glass slide covered with a thin layer of PDMS in order to have the same material on the four walls of the channel. The channels had a rectangular cross section and a constant height (either $H=57 \pm 3$ $\upmu$m or $H=103 \pm 3$ $\upmu$m). They were composed of three regions: two linear channels with different width (a large channel on the order of millimeters and a small channel on the order of a few hundreds of micrometers) were connected by a constriction with a small angle (5 to 10 degrees). \par

Slabs of hydrogel with controlled geometry and position were directly fabricated inside the micro-channel using microscope-based photolithography \cite{Dendukuri2008}. This method has been extensively detailed in previous works \cite{Dendukuri2008, Wexler2013, Berthet2016, Duprat2014, Nagel2018, Cappello2019, Bechert2019}, and just the essential steps are detailed here and schematically represented in  Figure~\ref{fig:schema_1}. The channel, filled with an oligomer solution and a photo-initiator, is exposed to a pulse of UV light of controlled duration. A mask with transparent drawings and black background is placed on the field-stop position of the microscope. 
The UV light passes through the mask and forms its image in the focal plane of the microscope. Thus, only the part of the channel corresponding to the transparent part of the drawings is illuminated by the UV light. Polymerization of the solution occurs in this region, allowing for the fabrication of hydrogel particles whose geometry corresponds to the drawings on the mask. After fabrication, the particles are surrounded by the uncured solution. The position of the particle only depends on the position of the channel relative to the objective of the microscope, which can be adjusted by moving the stage. Because the method of fabrication is based on a 2D projection technique, the 3D geometry of the particle results from the polymerization of the PEGDA in the height of the channel. The permeability of PDMS to oxygen (which inhibits the cross-linking reaction) leaves a non-polymerized lubrication layer of constant thickness on the top and bottom of the particles, allowing particles to be freely transported in the channel.

In this work, we used an inverted microscope (Zeiss Axio Observer) equipped with a UV light source (Lamp HBO 130 W) and a $\times 5$ Fluar objective.{ The shutter (Uniblitz, V25)), with 10 ms response time, was coupled with an external generator (Agilent 33220A) and allowed for a very accurate control of the exposure time}. We fabricated rectangular particles at zero flow rate in the wide region of the channel. The width $w_0$ and length $\ell_0$ were determined by the dimensions of the drawings on the mask corrected with a factor that accounts for the objective magnification. The particle height is $h_0=H-2b$ where $H$ is  the channel height and $b$ is the inhibition layer thickness (top and bottom), which we measured to be  $b=6 \pm1.6\, \upmu$m. Due to the dimensions of the microchannel, the height of the fabricated particle was always smaller than its width and length. The maximal dimensions of a particle are set by the UV light beam diameter, and was 1.5 mm in our setup. To ensure homogeneous crosslinking of the hydrogel we chose to limit the maximal dimensions of the particle to 1 mm.

 The photosensitive solution was composed of an oligomer, polyethylene glycol diacrylate (PEGDA, M$_{\rm n} = 700$ g/mol, Sigma), a photoinitiator (PI, 2-hydroxy-2-methylpropiophenone, Sigma), and a solvent. The solvent could be either pure water or a mixture of water and  polyethylene glycol (PEG$_{1000}$, M$_{\rm n} = 1000$ g/mol, Sigma) in a proportion of 1:2. While the proportion of solvent varied, the proportion of photointiator was kept constant at $10\%$. The volume fraction of water varied from $0\%$ to $50\%$. Above $50\%$ the solution became biphasic and no particle could be fabricated. The maximal dilution with the PEG$_{1000}$:water mixture was 70$\%$. Table~\ref{tab:oligomer_solution} shows the composition of the different solutions used in this study.

 \begin{table}[h]
\centering
\caption{Volume fraction of each component of the different  photosensitive solutions used in this study.}
\begin{tabular}{c c c p{3cm} c}
\toprule
  \multirow{ 2}{*}{\hfil PEGDA}	&  \multirow{ 2}{*}{\hfil PI } &  \multirow{ 2}{*} {\hfil Water}	& {\hfil PEG$_{1000}$:water} &  \multirow{ 2}{*}{\hfil Name}\\
 \hfil & \hfil  & \hfil  & \hfil {ratio 2:1 in volume} & \hfil  \\
\hline
\hfil 90$\%$		& \hfil 10$\%$  	&\hfil 0$\%$ 	& \hfil0$\%$ 			& \hfil pure PEGDA\\
\hfil 80$\%$		& \hfil 10$\%$  	&\hfil 10$\%$  	& \hfil 0$\%$ 			& \hfil PW$_{10}$\\
\hfil 70$\%$		& \hfil 10$\%$  	&\hfil 20$\%$  	& \hfil 0$\%$ 			& \hfil PW$_{20}$\\
\hfil 60$\%$		& \hfil 10$\%$ 	&\hfil 30$\%$   	& \hfil 0$\%$ 			&\hfil  PW$_{30}$\\
\hfil 50$\%$ 		& \hfil 10$\%$ 	&\hfil 40$\%$ 	& \hfil 0$\%$ 			& \hfil PW$_{40}$\\
\hfil 40$\%$		& \hfil 10$\%$ 	&\hfil 50$\%$ 	& \hfil 0$\%$ 			& \hfil PW$_{50}$\\
\hfil 80$\%$		& \hfil 10$\%$ 	&\hfil 0$\%$ 	& \hfil 10$\%$ 			& \hfil PP$_{10}$\\
\hfil 70$\%$		& \hfil 10$\%$ 	&\hfil 0$\%$ 	& \hfil 20$\%$ 			& \hfil PP$_{20}$\\
\hfil 60$\%$		& \hfil 10$\%$ 	&\hfil 0$\%$ 	& \hfil 30$\%$ 			& \hfil PP$_{30}$\\
\hfil 50$\%$		& \hfil 10$\%$ 	&\hfil 0$\%$ 	& \hfil 40$\%$ 			& \hfil PP$_{40}$\\
\hfil 40$\%$		& \hfil 10$\%$ 	&\hfil 0$\%$ 	& \hfil 50$\%$ 			& \hfil PP$_{50}$\\
\hfil 30$\%$		& \hfil 10$\%$ 	&\hfil 0$\%$ 	& \hfil 60$\%$ 			& \hfil PP$_{60}$\\
\hfil 20$\%$		& \hfil 10$\%$ 	&\hfil 0$\%$ 	& \hfil 70$\%$ 			& \hfil PP$_{70}$\\
\hline 
\end{tabular}
\label{tab:oligomer_solution}
\end{table}

  \subsection{Experimental Protocol}

\begin{figure*}
\centering
\includegraphics[width=0.9 \textwidth]{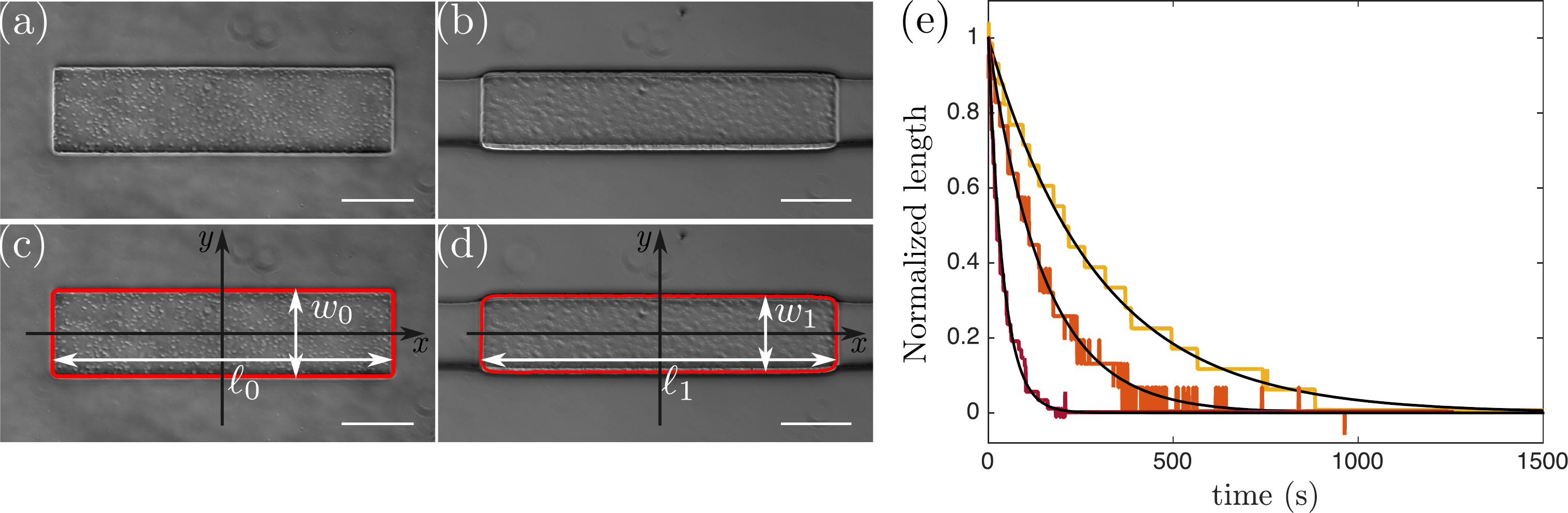}
\caption{Pictures of the slab in the wide {((\textbf{a}) and (\textbf{c}))}
~and in the narrow {((\textbf{b}) and (\textbf{d}))} {part of the} channel. {(\textbf{a}) and (\textbf{b}) correspond to raw images of the undeformed and deformed particle, respectively.} In (b) the dark region around the particle corresponds to the channel walls. {(\textbf{c}) and (\textbf{d}) show} the determination of the shape using {MatLab.} Scale bars are 200 $\upmu$m. (\textbf{e}) Temporal evolution of the particle length after the flow was stopped ($t=0$
). Length at time $t$ was normalized by subtracting the equilibrium length and dividing by the difference between the initial length and the final length. Three different hydrogels are represented:  
from  right to left, PP$_{20}$ (yellow), PP$_{30}$ (orange), and PP$_{40}$ (red). 
Exponential fits (black lines) are{ represented as a guide for the eyes.}} 
\label{fig:determination_interface}
\end{figure*}

The inlet of the micro-channel was connected to a reservoir, and a pressure controller (LineUP Series, Fluigent) was used to control the flow in the channel. Once the channel was filled with the photosensitive solution, the flow was stopped and a particle was fabricated. A picture of the particle was taken using a Hamamatsu Orca-flash 4.0 camera, which gave a precise measure of the dimensions before deformation (see Figure~\ref{fig:determination_interface}a). Then, the particle was pushed through the constriction into the narrow channel by imposing a flow through a pressure difference along the channel varying from a few hundred millibars to two bars. Once the particle had entirely entered the narrow region, the flow was turned off. There, the particle experienced uniaxial compression from the channel's lateral walls and we monitored its shape evolution with a frame rate of one image per second.
Using standard image treatment procedures (with ImageJ \cite{Schneider2012} and MatLab), we extracted the shape of the particle before and during deformation and measured the particle width and  length (see Figure \ref{fig:determination_interface}c,d). The length of the compressed particles decreased from its initial value to an equilibrium value (see Figure~\ref{fig:determination_interface}e). In all our experiments we made sure to wait long enough ($\sim 1200$s) to measure the length $\ell_1$ of the particle when this equilibrium was reached. At the end of the experiment, the particle was ejected from the narrow channel by again imposing {a pressure-driven flow}. We repeated this procedure at least 10 times for each condition.  In some cases, at the end of the experiment,  the particle was transported back into the wide channel where the particle shape was measured after the flow was stopped. We observed no residual deformation, confirming the reversibility of the deformation and ruling out any permanent deformations.
 
We varied the particle height ($h_0 = 45 \pm 3 \, \upmu$m and $h_0 = 91 \pm 3 \, \upmu$m) and  width ($w_0 =  205 - 470\,  \upmu$m) and the narrow channel width ($W = 175 - 370\, \upmu$m). We kept  the length of the particle constant and equal to $\ell_0 = 960\pm8 \,\upmu$m.  
 As can be seen in Figure~\ref{fig:determination_interface}b, the narrow channel was slightly deformed by the presence of the particle. We thus chose not to consider the width of the deformed particle to be equal to the width of the narrow channel but rather to measure $w_1$ for each experiment. The particle width $w_1$ differed only slightly from the channel width $W$ and the importance of the difference between the two measures is given by the ratio of the Young modulus of the PDMS (E$_{PDMS}$ $\sim$ 2 MPa \cite{Gervais2006}) and the hydrogel \cite{Cappello2019}. Particle deformation in the $y${-}direction will thus in the following be determined as $w_0/w_1$ instead of $w_0/W$.

 \subsection{Analysis}
 
After the full entry of the particle into the narrow channel, the external flow was stopped and the particle was only submitted to an uniaxial compression by the lateral walls. We assumed that the hydrogel was isotropic and homogeneous and linear elasticity thus gives: 

\begin{align}
\epsilon_{xx}  &=\frac{1}{E}[(1+\nu)\sigma_{xx}-\nu(\sigma_{xx}+\sigma_{yy}+\sigma_{zz})]{,} \label{eq:strain_stress_relation_x}\\
\epsilon_{yy}  &=\frac{1}{E}[(1+\nu)\sigma_{yy}-\nu(\sigma_{xx}+\sigma_{yy}+\sigma_{zz})]{,} 
\label{eq:strain_stress_relation_y}\\
\epsilon_{zz}  &=\frac{1}{E}[(1+\nu)\sigma_{zz}-\nu(\sigma_{xx}+\sigma_{yy}+\sigma_{zz})]{,} 
\label{eq:strain_stress_relation_z}
\end{align}

 with {\boldmath${\sigma}$ and \boldmath${\epsilon}$} respectively being the stress and strain {tensors} {of} the particle. {Equations (\ref{eq:strain_stress_relation_x}) to (\ref{eq:strain_stress_relation_z}) give the expression of the diagonal terms of \boldmath${\sigma}$ and \boldmath$\epsilon$.} $E$ is the Young modulus of the hydrogel and $\nu$ is its Poisson ratio.  
\par 
For the particle being submitted to a uniaxial compression in the $y$-direction, one can write $\sigma_{yy}=-P_{wall}$ and $\sigma_{xx}=0$. Assuming that the deformed particle does not touch the top and bottom walls (
{i.e.},  $h_1 < H$)  there is no stress in the $z$-direction  and $\sigma_{zz}=0$. This last assumption has to be verified {
a posteriori} by evaluating the strain in the $z$-direction.
 
 Equations (\ref{eq:strain_stress_relation_x}), (\ref{eq:strain_stress_relation_y}), and (\ref{eq:strain_stress_relation_z}) then become :

 \begin{align}
\epsilon_{xx}  =-\frac{\nu}{E} \sigma_{yy}, \,\, \, 
\epsilon_{yy}  =\frac{1}{E} \sigma_{yy}, \,\,\, \text{and}  \,\,\,
\epsilon_{zz}  =-\frac{\nu}{E} \sigma_{yy} .
\label{eq:strain_stress_relation_2}
\end{align}

{
The strains  $\epsilon_{xx}$  and $\epsilon_{yy}$ are related to the changes of particle length and width }
\begin{align}
\label{eq:strain_yy}
{\epsilon_{yy} =\ln \left( \frac{w_1}{w_0}\right), \,\, \, 
\epsilon_{xx} = \ln \left( \frac{\ell_1}{\ell_0}\right),}
\end{align}

{and Poisson's ratio is directly given by their ratio: }
\begin{align}
{\nu = -\frac{\epsilon_{xx}}{\epsilon_{yy}}=-\frac{\ln \left( {\ell_1}/{\ell_0}\right)}{\ln \left( {w_1}/{w_0}\right)}}.
\label{eq:Poisson_ratio}
\end{align}%

\section{{Results and Discussion}}

\subsection{Validation of the Method}

\begin{figure*}
\centering
\includegraphics[width=0.9 \textwidth]{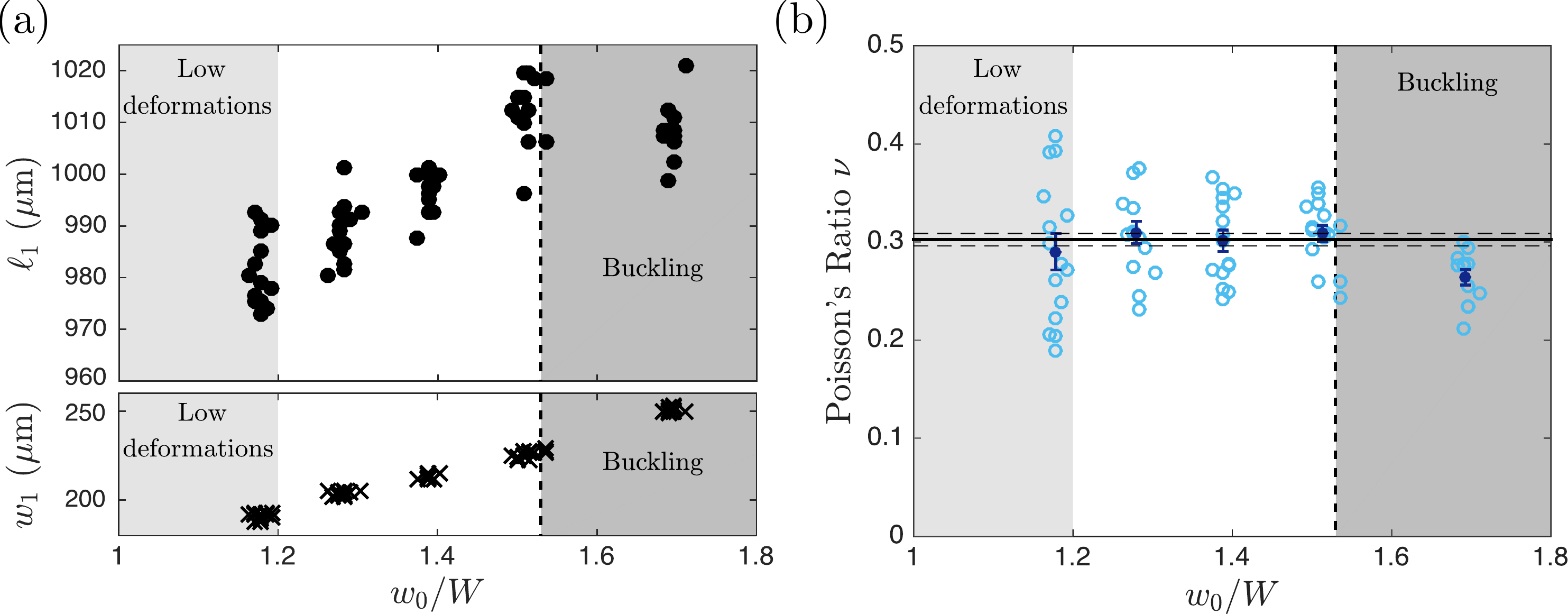}
\caption{Direct measurements of particle length and width and the derivation of Poisson's ratio. Results are shown for the hydrogel PW$_{30}$ in a channel of width $W=175\pm0.7\,\upmu$m and height $H=103 \pm 3 \, \upmu$m. The initial particle length was kept constant $l_0=960\pm 8 \,\upmu$m while the confinement $w_0/W$ was increased. (\textbf{a}) Length (top) and width (bottom) of the deformed particle at equilibrium as a function of the confinement $w_0/W$. (\textbf{b}) Evolution of Poisson's ratio as a function of $w_0/W$. {In (a) and (b) the} light (dark) gray regions correspond to particle widths that are too small (large) to ensure good precision of the determination of Poisson's ratio. The vertical dotted line corresponds to the buckling threshold (see text). In {(b)} each light-blue circle corresponds to a single experimental measurement, and dark blue markers correspond to their average and standard {deviation $\sigma$ normalized by $\sqrt{n}$ with $n=15$ independent measurements}. The horizontal black lines correspond to the average {(solid line)} and standard deviation {(dashed line)} of all points except from the dark gray region ($w_0/w_1\geq  1.53$).
Poisson's ratio was found to be $\nu_{30\%} = 0.
302 \pm 0.007$ for the photosensitive solution PW$_{30}$.}
\label{fig:donnes_brutes}
\end{figure*}

When the particle was compressed by the channel walls its length increased, as can be seen in Figure~\ref{fig:determination_interface}a--d, as expected for an elastic material. 
Figure~\ref{fig:donnes_brutes} summarizes experiments where particles were created in a PW$_{30}$ solution and  compressed into a channel of constant width $W=175\pm0.7\,\upmu$m. The initial particle width $w_0$ was varied to apply varying confinements $w_0/W$. To extract Poisson's ratio we computed the ratio  $-\epsilon_{xx}/\epsilon_{yy}$  according to Equations (\ref{eq:Poisson_ratio}) taking into account that the width of the deformed particles was slightly different from the width $W$ of the channel as a consequence of the deformation of the lateral channel walls (bottom of Figure \ref{fig:donnes_brutes}a).
Figure \ref{fig:donnes_brutes}b summarizes our measurements of $\nu$ for different values of the confinement $w_0/W$. Each data point (blue circles) corresponds to one compressed particle and the average values and corresponding error bars are represented in {dark blue}. 
Larger $w_0/W$ induces stronger deformation and thus larger stresses applied by the walls on the particle. In Figure \ref{fig:donnes_brutes}b, Poisson's ratio is independent of $w_0/W${ in the range [1 1.53]} {corresponding to the applied strain varying from $-6.2\%$ to $-17.7\%$}, validating our technique to measure a material property which is expected to be independent of the geometry, the applied stress, and the resulting strain. For very small ratios $w_0/W$ ($w_0/W \leq 1.2$, light gray region) particle deformation was small and large scatter of the data points was observed. For the widest particle ($w_0=300\upmu$m), the stress was large enough to induce buckling (see the images in {Figures~\ref{fig:std_evolution}a,b}), leading to a saturation of $\ell_1$ and preventing any measurement of Poisson's ratio.

\subsection{Limitations}

Limitations are inherent to every technique and it is essential to take them into consideration when developing a new method. In the following {section}, we discuss the range of validity for our microfluidic technique. First, our analysis relies on the assumption that the compressive stress was solely applied in the $y$-direction by the lateral walls of the narrow channel and thus the only stress component that was {non-zero} was $\sigma_{yy}$. For this to be true deformation in the {vertical direction} needs to be small enough so that the particle does not touch the channel top and bottom walls, which would lead to an additional compressive stress $\sigma_{zz}$ in the $z$-direction. This can be verified by comparing the height of the compressed particle to the channel height. The height of the deformed particle is $h_1 =  h_0 \exp(\epsilon_{zz}) \sim h_0(1-\nu \epsilon_{yy})$ and the condition $\sigma_{zz} = 0$ is verified as long as $- h_0 \nu \epsilon_{yy} < 2b$. As a consequence the assumption is more likely to be verified for particles of small height $h_0$ and for small deformations. In all our experiments we verified 
{a posteriori} that the deformed particle height remained smaller than the channel height and thus that the assumption  $\sigma_{zz} = 0$ was valid. We disregarded experiments where this was not the case. Note also that the condition $\sigma_{xx}=0$ was not strictly verified close to the edges of the hydrogel particle due to the slight deformation of the PDMS channel  (see Figure~\ref{fig:determination_interface}b,d). This effect is more important for particles of large elastic modulus and strong compression $\epsilon_{yy}$, but is considered to be a small correction for our experiments.

Second, for too-small confinements  $w_0/W$ particle deformation was small and the experimental accuracy of $\nu$ was limited by the resolution of the deformed particle dimensions. The resolution could be improved  either by applying  stronger confinement (increasing the ratio $w_0/W$) or by upscaling the experiment, increasing both $w_0$ and $W$. However, we will see below that this can  favor buckling instabilities and is thus not a favorable option in all cases.

\begin{figure*}[ht]
\centering
\includegraphics[width=0.9 \textwidth]{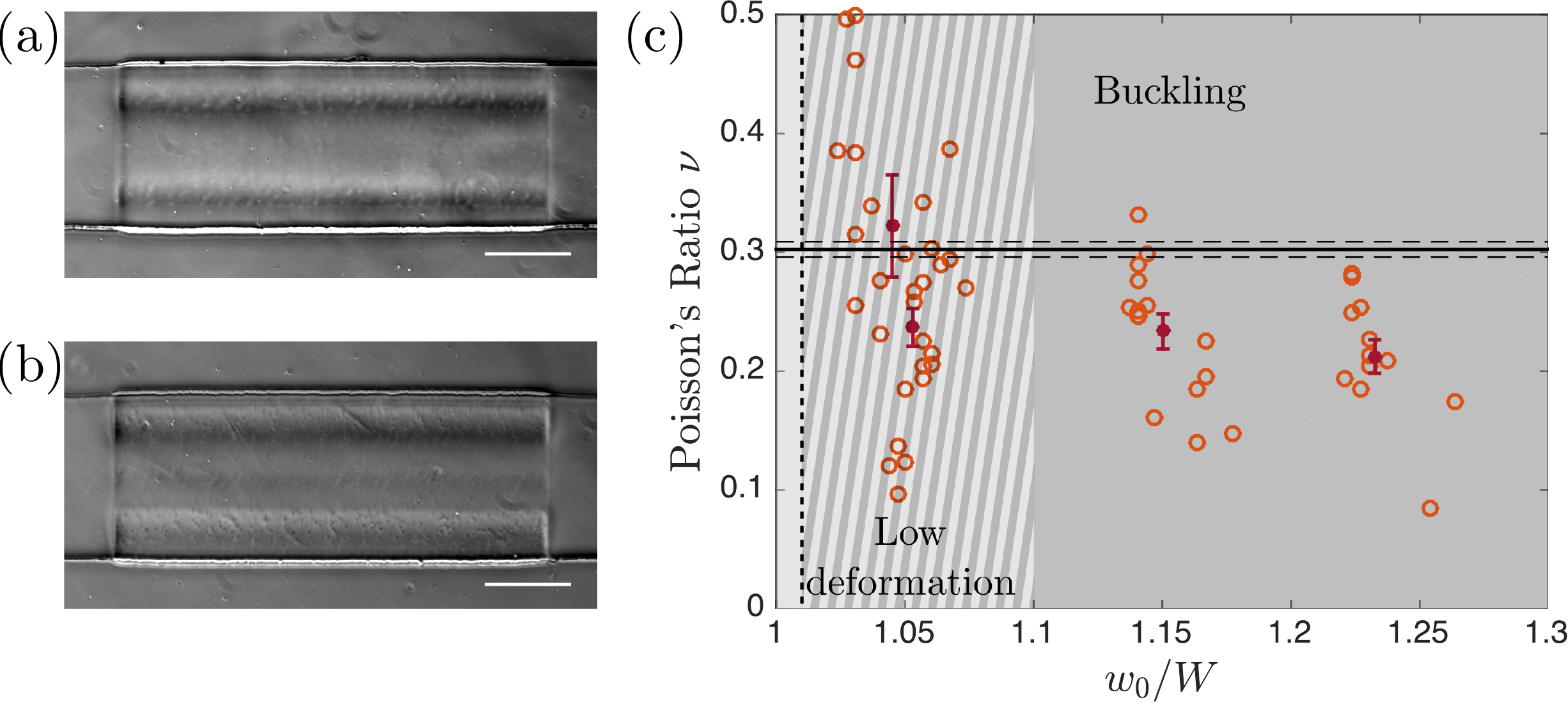}
\caption{Limitations of the method.  Results correspond to experiments performed with hydrogel particles created with the PW$_{30}$ photosensitive solution in a channel of width $W=370 \pm 1\,\upmu$m  and height $H= 57 \pm 3 \,\upmu$m. (\textbf{a}) and (\textbf{b}) show two pictures of buckled particles, at least two wavelengths are visible in (a) and three in (b). They correspond respectively to $w_0/W = 1.23$ and $w_0/W = 1.36$. (\textbf{c}) Evolution of the ratio $-\epsilon_{xx}/\epsilon_{yy}$ as a function of $w_0/W$. Open circles represent the results of single experiments and dark red markers represent average values of the different experiments. The light/{dark} gray regions of Figure~\ref{fig:donnes_brutes}b overlap for this example. Note that here also the measured ratio $-\epsilon_{xx}/\epsilon_{yy}$ is lower than Poisson's ratio when the particle buckles.}
\label{fig:std_evolution}
\end{figure*}

Third, when the slab is too wide buckling may occur, preventing the measurement of Poisson's ratio.
The critical stress to induce buckling in a thick elastic plate is given by $\sigma_{yy}^{crit} = \alpha \pi^2\frac{E}{12(1-\nu^2)}\left(\frac{h_0}{w_0}\right)^2$ \cite{Jones2006, Onya2018}. The correction factor $\alpha$ depends on the ratio $w_0/h_0$ and takes into account the shear deformation in the height. According to \cite{Onya2018}{,} $\alpha=2.11$ for $w_0/h_0 = 2$  and $\alpha=1.03$ for $w_0/h_0 = 10$. This critical stress is compared to the stress exerted by the lateral walls assuming linear deformation, $\sigma_{yy}= \epsilon_{yy} E$, via the dimensionless number 
 \begin{align}
 N=\frac{12(1-\nu^2)}{\alpha \pi^2}\left(\frac{w_0}{h_0}\right)^2\epsilon_{yy}.
 \label{eq:buckling}
 \end{align}
 If $N>1$ the particle buckles and, on the contrary, if $N<1$ the particle deforms linearly. The black vertical dotted line in Figure \ref{fig:donnes_brutes} corresponds to $N=1$ and good agreement between the theoretically predicted buckling threshold and the experimental observations was obtained. Buckling of the hydrogel particle was visible by direct observation (see Figures \ref{fig:std_evolution}a,b) but also from the saturation of $\ell_1$ at strong confinement (see Figure~\ref{fig:donnes_brutes}a) associated to the non-physical decrease of $\nu$ seen in Figure~\ref{fig:donnes_brutes}b. Since the buckling threshold $N$ depends on Poisson's ratio as $1-\nu^2$, variations of $\nu$ between 0 and 0.5 induce only very small modifications of $N$. Thus, the buckling threshold is mainly given by the geometry of the particle (via the particle aspect ratio $w_0/h_0$) and the geometry of the channel (the strain $\epsilon_{yy}$ being proportional to $w_0/W$). In conclusion, to avoid buckling, small $\epsilon_{yy} \propto w_0/W$ and small aspect ratios ($w_0/h_0$) are favorable.

To illustrate these  findings, in Figure \ref{fig:std_evolution}c both the channel width and height were modified compared to Figure \ref{fig:donnes_brutes} while keeping the same photosensitive solution. The channel width $W$ was increased  to have a better resolution and the limit of the light grey region was indeed pushed to a lower value of $w_0/W$ ($\approx 1.1$ to be compared to $\approx 1.2$ in Figure \ref{fig:donnes_brutes}b). The channel width on the other side was increased and the particles were thus more prone to buckle. The limit of the dark grey region was shifted to smaller values of $w_0/W$ ($\approx 1.01$ compared to $\approx 1.53$ in Figure \ref{fig:donnes_brutes}b). As can be seen in the figure, despite the better resolution, the two grey regions overlap, meaning that this channel geometry does not allow for Poisson's ratio measurements.

 The comparison of Figures \ref{fig:donnes_brutes}b and \ref{fig:std_evolution}c illustrates the compromise that has to be made between increasing the resolution on the particle deformation and avoiding particle buckling. A decrease of $w_0$, keeping $h_0$ constant, increases the range of deformation before buckling but at the same time  decreases the resolution. On the other hand, an increase of $h_0$ alone is favorable to avoid buckling but should be limited to prevent the deformed particle from touching the top and bottom walls. While designing channels for such measurements, one has to keep these two opposite effects in mind and to choose the best geometry varying the channel width and height accordingly.
 
 Finally, for very soft particles and thus weakly crosslinked hydrogel particles ({PP$_{60}$ and PP$_{70}$} \cite{Cappello2019}), we observed a dependency of the equilibrium length ($\ell_1$) on the velocity at which a particle enters into the constriction. A possible reason for this observation can be found in the viscoelastic properties of the gel and/or the complex friction between the hydrogel and the lateral channel walls \cite{Gong2006}. In such situations we consider that the measurement of Poisson's ratio cannot been achieved properly.



\subsection{Dependence of Poisson's Ratio on the Solvent Composition}

  \begin{figure*}[ht]
\centering
\includegraphics[width=0.6 \textwidth]{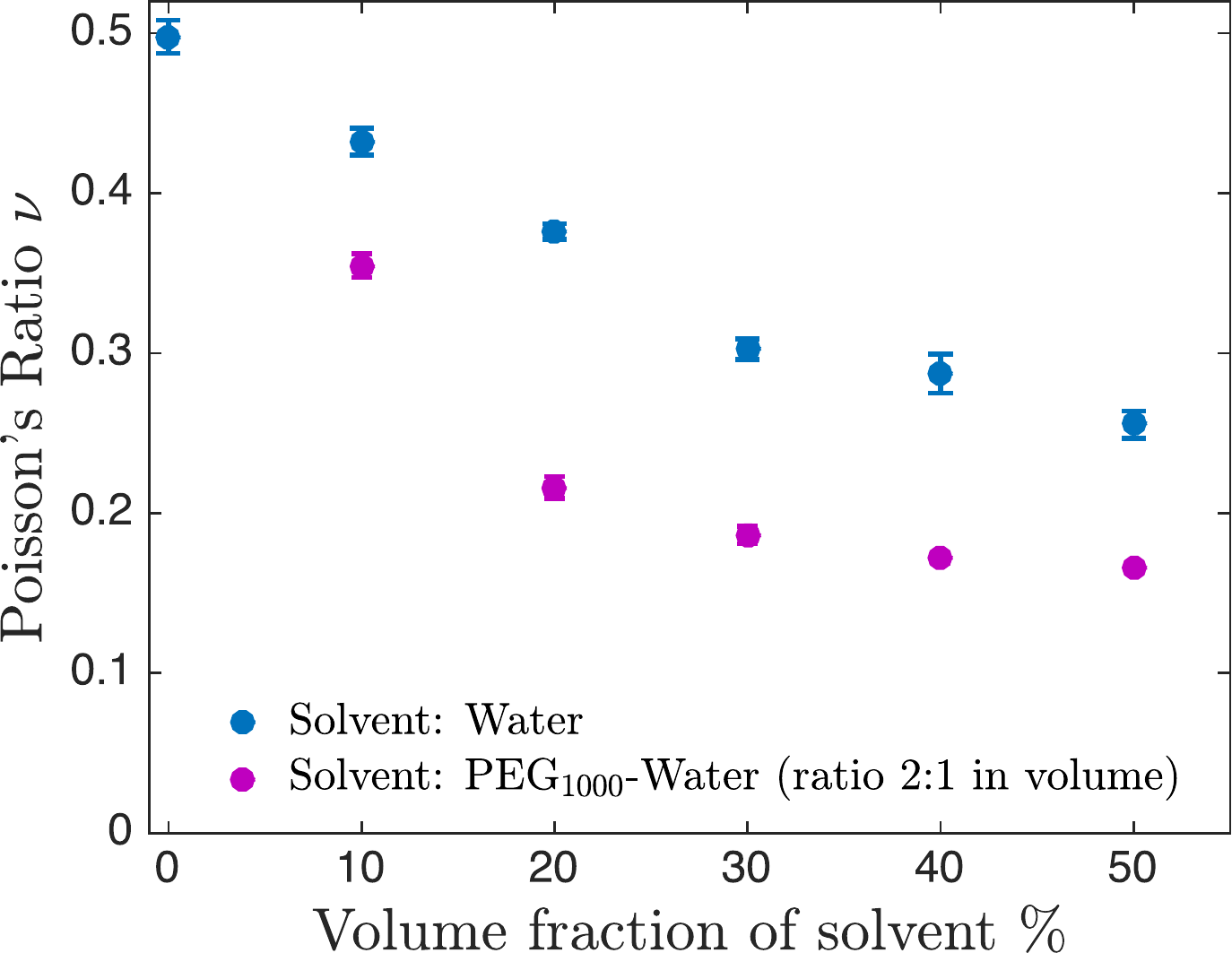}
\caption{ Evolution of the Poisson ratio of hydrogels fabricated from different photosensitive solutions as a function of the solvent volume fraction. Two different solvents are shown: water (blue markers) and a solution of PEG M$_{\rm n}=1000$ g/mol:water at a volume ratio of 2:1 (purple markers).}
\label{fig:determination_poisson}
\end{figure*}

Using our method we measured {Poisson's ratio} of hydrogels fabricated from different compositions of photosensitive solutions with different solvent concentrations and solvent nature. Figure~\ref{fig:determination_poisson} shows the measured Poisson's ratio for dilutions of PEGDA M$_{\rm n}=700$ g/mol with water (blue points) and with a PEG$_{1000}$:water mixture (2:1 in volume, purple points). In the absence of solvent the hydrogel was nearly incompressible ($\nu_{0\%} = 0.50 \pm 0.01$) which is in agreement with the literature \cite{Gabler2009}. In the presence of solvent, the hydrogel became compressible, as shown by the decrease of Poisson's ratio. This compressibility results from the flow of solvent molecules leaving the particle when the hydrogel is compressed and depends on the microstructure of the cross-linked polymeric chains of PEGDA. Note that the hydrogel particles are incompressible during an isotropic compression  imposed by a static pressure, as a consequence of the incompressibility of water. For larger dilution of the photosensitive solution, the fraction of solvent increases, leading to a smaller Poisson's ratio. In addition, we observed that for the same dilution, Poisson's ratio was always smaller in PEG$_{1000}$:water compared to pure water. In water, the lowest value measured was $\nu_{50\%}=0.255\pm0.009$ while in PEG$_{1000}$:water this value was $\nu_{50\%}=0.165 \pm 0.002$. The decreased value of Poisson's ratio reported here is a signature of the modified microstructure of the hydrogels. Further work is however necessary to fully understand the impact of the presence of chains of PEG in the surrounding medium on the compressibility of the hydrogels. Several hypotheses can be proposed: first, the presence of PEG is known to increase the porosity of the hydrogel \cite{Lee2010}, which is expected to modify its mechanical properties. The presence of PEG chains probably also impacts the connectivity of the meshwork as well as the chemical potentials of the different species. \par

The flow of solvent through the polymeric mesh induced poroelastic effects, explaining the temporal evolution of the particle length shown in Figure~\ref{fig:determination_interface}e. Poroelasticity describes the flow of solvent through the mesh of a deformable material and introduces a time-dependent response to deformation. According to references  \cite{Cai2010} and \cite{ Hu2010} this poroelastic phenomenon is expected to depend on the porosity of the hydrogel and the viscosity of the solvent. The three different photosensitive solutions used in Figure~\ref{fig:determination_interface}e, PP$_{20}$ (yellow), PP$_{30}$ (orange), and PP$_{40}$ (red), have comparable viscosities ($\mu^{PP_{20}}=108\pm3$ mPa$\cdot$s, $\mu^{PP_{30}}=107\pm3$ mPa$\cdot$s, and $\mu^{PP_{40}}=116\pm3$ mPa$\cdot$s). On the contrary, the mesh size decreased with increasing dilution as the polymer chains were further apart during crosslinking in the presence of solvent molecules. This is in good agreement with the tendency shown in Figure~\ref{fig:determination_interface}e in which the characteristic {relaxation} time decreased when the solvent fraction increased. Note that as the poroelastic timescale also depends on the dimensions of the hydrogel, the timescales for  micron-scale particles as in the present study are much smaller than for macroscopic particles \cite{Cai2010, Hu2010}. Let us state again that all our measurements were performed at equilibrium to prevent any impact of these temporal effects on our measurement of the hydrogel compressibility.

\section{Conclusion} 

\begin{figure*}[ht]
\centering
\includegraphics[width=0.7 \textwidth]{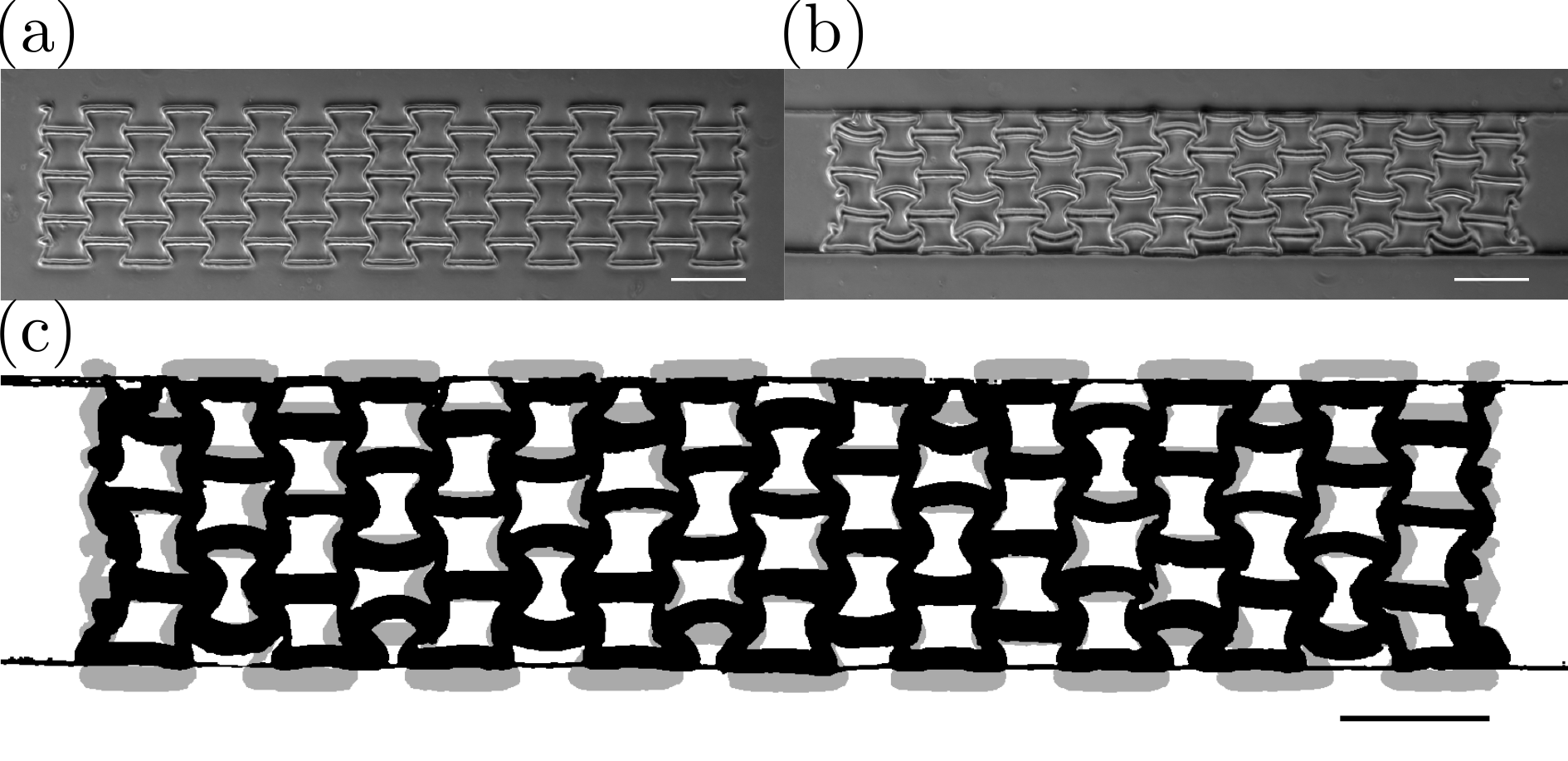}
\caption{Auxetic metamaterial fabricated from a solution of 90$\%$ PEG-DA M$_{\rm n}=700$ g/mol and 10$\%$ photo-initiator in volume. (\textbf{a}) Before compression, (\textbf{b}) submitted to an uniaxial compression, (\textbf{c}) superposition of the undeformed and deformed shapes. The particle width  and length decreased when compressed, {which is indicative}
~of a negative Poisson's ratio.}
\label{fig:poisson_negatif}
\end{figure*}

Until now we have considered homogeneous materials and shown that our experimental technique is well-suited to determine the Poisson ratio of the material they are made of. Figure~\ref{fig:poisson_negatif} shows the extension of our technique to metamaterials in which the geometry of the structure changed the mechanical properties. In this example, an auxetic particle was fabricated using a specific design of the structure \cite{Ren2018}. The superposition of the shapes of this particle before (gray) and after (black) introduction into the narrow channel clearly shows that the particle was slightly shorter when compressed, which is the signature of a negative Poisson's ratio.

We  presented a new technique for the measurement of {Poisson's ratio} of micrometric soft hydrogels{ with a very good measurement accuracy. The absolute measurement errors varying from 0.002 to 0.012 are comparable to what is obtainable using X-ray diffraction \cite{Moram2007}, and are much smaller than for measurements relying on the determination of two independent elastic moduli \cite{Punter2020, Wyss2010}}.  We successfully used this approach to measure {Poisson's ratios} of different hydrogels and showed that this Poisson ratio varied in a large range of values (0.165 to 0.5). We discussed the limitations of the methods and showed how channel and particle geometries should be chosen to accurately measure Poisson's {ratio}. Compared to other approaches, an important advantage of our method is that it is done {
in situ} directly in the surrounding fluid and does not require any drying of the sample. Another advantage as compared to atomic force microscopy (AFM), which probes the mechanical properties at the nanometer scale, is the determination of the mechanical property (Poisson's ratio in our case) of the entire hydrogel particle {on a scale much larger than the pore sizes even for very porous materials}. Moreover, the technique presented here allows for an excellent measurement of the dynamical response of the material, enabling the determination of Poisson's ratio at long times and thus at equilibrium. Our characterization method is not limited to particles directly fabricated inside a microchannel but can also be used, for example, for protein or cell aggregates that can be flown into the microchannel. However, their potentially more complex shape and porosity can make the analysis less {straightforward} \cite{Duchene2020}{ and would require the use of effective strains and stresses to describe the particle deformation \cite{Bommireddy2019}. Moreover, if the condition $\sigma_{zz}=0$ is not verified, our method could be adapted to measure Poisson's ratio, which in that case derives from the formula $\nu/(1-\nu) = - \epsilon_{xx}/(\epsilon_{yy}+\epsilon_{zz})$.} Because of these advantages and its simplicity, we believe that this method will be used for the characterization of {Poisson's ratio} of many different soft objects from nature and industry. 

\vspace{6pt} 

\begin{acknowledgments}
{We thank Al Crosby for inspiring discussions and Charles Duch\^ene for help with the experimental set-up.} \\ \\
{This research was funded by the European Research Council through a consolidator grant (ERC PaDyFlow 682367). This work received the support of Institut Pierre-Gilles de Gennes (\'Equipement d'Excellence, ``Investissements d'Avenir", Program ANR-10-EQPX-34). }
\end{acknowledgments}



\end{document}